\documentclass{article}

\usepackage[a4paper]{geometry}

\usepackage{amsmath}
\usepackage{graphicx}
\usepackage{caption}
\usepackage{subcaption}
\usepackage[hidelinks]{hyperref}

\newcommand{\relBone}{k_8}
\newcommand{\diff}{k_5}
\newcommand{\genBioB}{k_{2,1}}
\newcommand{\decBioB}{k_{3,1}}
\newcommand{\genBioV}{k_{2,2}}
\newcommand{\decBioV}{k_{3,2}}
\newcommand{\genV}{k_6}
\newcommand{\proV}{k_7}
\newcommand{\genB}{k_4}
\newcommand{\strainSense}{f}
\newcommand{\vascSpace}{\varphi}
\newcommand{\cOst}{c_{\textrm{ost}}}
\newcommand{\cVasc}{c_{\textrm{vasc}}}
\newcommand{\aBone}{a_{\textrm{BGF}}}
\newcommand{\aVasc}{a_{\textrm{VEGF}}}
\newcommand{\decay}{k_1}

\begin{document}
\begin{center}
{\Large
A parameter study on optimal scaffolds\\ in a simple model for bone regeneration\\[5mm]}
{\today}\\[5mm]
 {Patrick Dondl, Marius Zeinhofer}\\[2mm]
{\em Abteilung für Angewandte Mathematik,
 Albert-Ludwigs-Universität Freiburg, \\
 79104 Freiburg i. Br., Germany}
\end{center}

\begin{abstract}
We propose a simple model for scaffold aided bone regeneration. In this model, only macroscopic quantities, e.g., locally averaged osteoblast densities, are considered. This allows for use of this model in an optimization algorithm, whose outcome is an optimal scaffold porosity distribution. This optimal scaffold naturally depends on the choice of parameters in the model, and we provide a parameter study with a particular focus on patients with reduced bone regeneration or reduced vascularization capacity.
\end{abstract}

\section{Introduction}
We consider a model for bone regeneration in the presence of a resorbable polymer scaffold and use this model to derive scaffold designs with optimal properties for bone tissue engineering. Particular focus here lies on the treatment of patients with critical size bone defects \cite{schemitsch2017size} who are in addition suffering from a co-morbidity that may lead to a reduced bone regeneration capacity, e.g., type 2 diabetes mellitus (T2DM). This metabolic disease for example leads to reduced vasculature \cite{alcaraz2017novo} and innervation \cite{alcaraz2014keypathwayminer}, as well as a reduced rate of formation of bone matrix \cite{list2016keypathwayminerweb}. Overall, the treatment of non-unions associated with critical size bone defects in the presence of metabolic disease are a difficult clinical problem \cite{alcaraz2017novo}.

The use of additively manufactured polymer scaffolds in bone tissue engineering has shown promising results \cite{Berner:2013,Cipitria:2013,Reichert:2011,Sawyer:2009,Rai:2007,Ip:2007,Hoda:2016,Teo:2015fu,Schuckert:2008hg,reznikov2019individual}. When employing current additive manufacturing techniques, a great freedom of design is afforded, so that scaffolds can be specifically designed in accordance with an individual patient's requirements, depending on, e.g., metabolic markers, defect site and geometry, and expected mechanical loads \cite{Poh:2016cu}. This can ensure that the often competing design objectives of facilitating tissue in-growth and providing mechanical stability are both met in a satisfactory manner.

So far, the design of scaffolds for bone tissue engineering has largely been based on a trial-and-error approach, but more recently topology optimization \cite{dondl_topopt, Adachi06, COELHO2015287, DIAS2014448} and other optimization methods \cite{dondl_opt} have been proposed. In this work, we follow the latter approach, which is based on homogenized quantities (also used in \cite{dondl_3d}). While models for bone regeneration in the presence of scaffolds that resolve the microscopic (scaffold microstructure and even cellular) scale exist \cite{perier2020mechano, CHECA2010961, Sanz-Herrera.2008, perez1}, they are generally difficult to employ in optimization algorithms, due to their high numerical complexity. Here we thus make two simplifying assumptions: first, that bone regeneration can be sufficiently accurately predicted using only quantities averaged over local representative volumes, i.e., macroscopic quantities. Second, we assume that there is one prominent axis (e.g., a principle stress axis in a long bone under compression) so that a one-dimensional model (using quantities averaged over slices) may be used.

The local scaffold density (or, equivalent, the scaffold porosity as one minus density), averaged over a suitable representative volume (e.g., a unit cell in the case of a periodic scaffold geometry) is now an input to such a model of bone regeneration. The solution of the model, i.e., the regenerated bone density, will depend on this input. Using a suitable objective function depending on the  model solution, one can now find an \emph{optimal} scaffold density.

In this work, we apply the adjoint approach of PDE constrained optimization \cite{hinze} to determine the sensitivity of optimal scaffolds on various parameters in the the model, in particular the rate of bone regeneration, the rate of vascularization, and the mechanical stiffness of regenerated bone matrix. As noted above, the first two of these parameters may be reduced, e.g., in T2DM patients, the third parameter may be reduced, e.g., in patients suffering from osteoporosis.

The remainder of this article is organized as follows. In section \ref{sec:model} we briefly introduce our model and the optimization objective for bone regeneration. Section \ref{sec:results} shows the outcome of our optimization for different choices of parameters.

\section{A model for bone regeneration and optimization of scaffolds} \label{sec:model}
Our model is a 1d, homogenized, simple model that is an extension of the one proposed in \cite{dondl_opt, dondl_3d}. As emphasized there, the model is designed to allow scaffold architecture optimization. Special focus lies on the role of (in T2DM patients suppressed) vascularization and bone regeneration and the implications thereof to the optimal scaffold design.

The model takes into account the time aspect of tissue engineering, in particular, bone growth, scaffold degradation and the interaction of the two. This allows to track the time-dependent mechanical environment of the scaffold-bone composite. The relevant quantities are the scaffold's local volume fraction $\rho(x)$, the molecular weight of the scaffold material $\sigma(t) = \exp(-\decay t)$ (which diminishes exponentialy over time due to bulk erosion) and the local volume fraction of regenerated osteoblast cells $\cOst$ that contribute with the mechanical properties of calcified bone, see \cite{perier2020mechano}. Bone regeneration, i.e.,\ the growth of osteoblasts, depends on the local biological environment modeled through growth factors/cytokines. Clinically, numerous such factors can be observed \cite{devescovi2008growth}, however, having vascularization in focus, we include only vascular endothelial growth factor (VEGF) -- responsible for new vessel formation -- and bone growth factor (BGF) which drives bone growth. These quantities are represented as $\aVasc(t,x)$ and $\aBone(t,x)$. Finally, $\cVasc(t,x)$ is the local fraction of endothelial cells responsible for vascularization.

In our approach, scaffold-mediated bone growth is modeled through a coupled system of evolution equations that describes the relationships between the aforementioned quantities. Constants in the model are either obtained from experimental observations or used in the sensitivity analysis presented in this manuscript. As mentioned before, all quantities appearing in the model can be understood as meso-scale averages over a number of pores in the material and are furthermore normalized to unity. See \cite{dondl_3d} for an illustration.

The spatial domain of computation is the space occupied by the scaffold, which is simplified to a one-dimensional object via considering only the main stress axis in a segmental defect in a long bone. More precisely, the defect is assumed to be $30$mm in size, which is resolved by the domain $\Omega = (0,30)$. The time horizon is set to $12$ months using the time interval $I=[0,12]$. Concretly, we solve the following system of differential equations
\begin{align}
    0 &= \operatorname{div}\left( (k_8 o + \rho\sigma)u' \right) \label{eq:elasticity}
    \\
    \dot{a}_1 &= \operatorname{div}\left( k_5(1-\rho)a_1' \right) + k_{2,1}f(u')o - k_{3,1}a_1 \label{eq:BGF}
    \\
    \dot{a}_2 &= \operatorname{div}\left( k_5(1-\rho)a_2' \right) + k_{2,2}o - k_{3,2}a_2 \label{eq:VEGF}
    \\
    \dot{v} &= k_6a_2(1+k_7v)\left( 1 - \frac{v}{1-\varphi(\rho)} \right)\label{eq:vascularization}
    \\
    \dot{o} &= k_4a_1v\left( 1 - \frac{o}{1 - \rho} \right)\label{eq:osteoblast}
\end{align}
In this system, $k_i$, $i=1, \dots, 8$ are parameters and $\strainSense$, $\vascSpace$ are functional relationships. Equation \eqref{eq:elasticity} allows to compute the displacement field $u(t,x)$ depending on the scaffold-bone composite. In equation \eqref{eq:BGF}, the term $\genBioB\strainSense(u')\cOst$ encodes that BGF gets only produced if osteoblasts sense a suitable mechanical stimulus. Furthermore, the BGF molecules diffuse and decay at certain rates. A similar behavior is modeled for VEGF, however, the production of VEGF does not require the presence of mechanical stimuli. The cell types responsible for vascularization, $\cVasc$ and for bone growth $\cOst$ are modeled as logistic ODEs pointwise in space. We do not include diffusion as these cells diffuse little if at all \cite{perier2020mechano}. The growth of $\cVasc$ is driven through the presence of $\aVasc$ and proliferation and is saturated by the ``space-filling'' factor $1-(\cVasc)/(1-\vascSpace(\rho))$. The functional relationship $\vascSpace$ represents the special need of blood vessels for space and should be chosen to lie above the identity function. It is now for example the sensitivity of the optimal scaffold design with respect to the parameters $\genV$ and $\proV$ that we focus on in this manuscript. Finally, equation \eqref{eq:osteoblast} for osteoblast production is similar to the one previously described. The necessary drivers are here BGF and $\cVasc$.

The initial and boundary conditions are given by 
\begin{align}
    u'(t,0) &= -u'(t,x) = 0.01 \label{eq:stress}
    \\
    \aVasc(0,x) &= \aBone(0,x) = 0
    \\
    \aVasc'(t,0) &= \aVasc'(t,L) = 0
    \\
    \aBone(t,0) &= \aBone(t,L) = 1
    \\
    \cVasc(0,x) &= 0
    \\
    \cOst(0,x) &= 0, \label{eq:bcost}
\end{align}
for all $x\in (0,L)$, $t\in [0,T)$, meaning that the elastic equation is subjected to a soft compressive load, BGF diffuses from healthy bone and VEGF is subjected to non-flux boundary conditions. Both molecules are not present at the initial time point. Similarly, neither $\cVasc$ nor $\cOst$ is present in the beginning after implantation of the scaffold.

The function $\strainSense$ is simply a regularization of the usual absolute value, with $\strainSense(u') = \sqrt{u'^2 + \delta^2}$, where $\delta = 10^{-4}$. The special cut-off function $\vascSpace$ is chosen as $\vascSpace(\rho) = \sqrt{\rho}$. This ensures that vasculature only occurs to the extent that there is space in the scaffold pores.
The default model parameters follow to a large extent the ones in \cite{dondl_3d} and are reported in Table \ref{tab:parameters}. They are chosen such  that the model with default parameters reproduces the results from the ovine model in \cite{Cipitria:2015kp}, see Figure \ref{fig:default}.

\begin{table}
\centering
\begin{tabular}{c|c||c|c||c|c||c|c||c|c}
$\decay$ & 0.1 & $\relBone$ & 9.0 & $\diff$ & 260 & $\genBioB$ & 12\,000 & $\decBioB$ & 16.0 \\
\hline
$\genBioV$ & 8.0 & $\decBioV$ & 8.0 & $\genV$ & 0.8 & $\proV$ & 1.2 & $\genB$ & 1.2 \\
\end{tabular}
\caption{Parameters in the default model.} \label{tab:parameters}
\end{table}

The proposed model can now be used to evaluate the time evolution of the mechanical stiffness of the implant for a given $\rho$. In our case this is proportional to the inverse of the elastic energy in the system at time $t\in[0,T]$
\begin{equation}
    E^{\textrm{el}}_\rho(t)^{-1} = \left[ \int_0^L u'(\rho\sigma + \relBone\cOst)u'\mathrm dx\right]^{-1}
\end{equation}
where $\rho$, $\sigma$, $u$ and $\cOst$ solve equations \eqref{eq:elasticity}--\eqref{eq:osteoblast} with conditions \eqref{eq:stress}--\eqref{eq:bcost}. Then we propose, similar to \cite{dondl_opt}, the objective function 
\begin{equation}
    J(\rho) =  \left(\int_0^T |E^{\textrm{el}}_\rho(t)|^5\,\mathrm{d}t\right)^{\frac{1}{5}},
\end{equation}
which is a good approximation for $\max_{t\in [0,T]} E^{\textrm{el}}_\rho(t)$, and the optimization problem becomes
\begin{equation}\label{eq:optimizationP}
    \min_{\rho}J(\rho) \quad \text{subjected to} \quad 0 < c \leq \rho(x) \leq C < 1,
\end{equation}
where the pointwise constraint on $\rho$ assures that it is a reasonable volume fraction. Note that the problem of minimizing $J$ is a PDE constrained optimization problem, as implicitly the definition of $J$ requires that the the system \eqref{eq:elasticity}--\eqref{eq:osteoblast} of partial differential equations with conditions \eqref{eq:stress}--\eqref{eq:bcost} is solved when used to compute $E^{\textrm{el}}$. A suitable technique to tackle problems of this type is the adjoint method in PDE constrained optimization, see for instance \cite{hinze}, which is also the approach we took. The cutoff in \eqref{eq:optimizationP} at $C=0.4$ has been implemented as a differentiable penalty function, again equivalent to the approach used in \cite{dondl_opt}. This cutoff must be introduced to prevent scaffold designs with too low porosity $1- \rho$. The lower bound $c$ is not relevant in practice, as too high porosities automatically yield unfavourable values for the objective functions. 

The main question we address in this manuscript is dependence of the optimal scaffold architecture on the parameters $\genB$, $\genV$, and $\proV$ -- that means, we solve the optimization problem for different values of these parameters and observe the optimiztion outcome.

The numerical implementation of the model is based on a simple semi-implicit in time (in the sense that equations \eqref{eq:elasticity}--\eqref{eq:osteoblast}, each of which is a linear equation, are solved individually implicitly one at a time in order) one-dimensional finite element scheme with 100 one-dimensional P1 finite elements and a time discretization  using 500 time steps for the interval $[0,T]$. The optimization problem is solved using an $L^2(\Omega)$ gradient flow with the variation of the objective function computed using an adjoint approach.

\section{Results and discussion} \label{sec:results}
\begin{figure}
\centering
\begin{subfigure}[t]{0.3\textwidth}
\centering
\includegraphics[width=\textwidth]{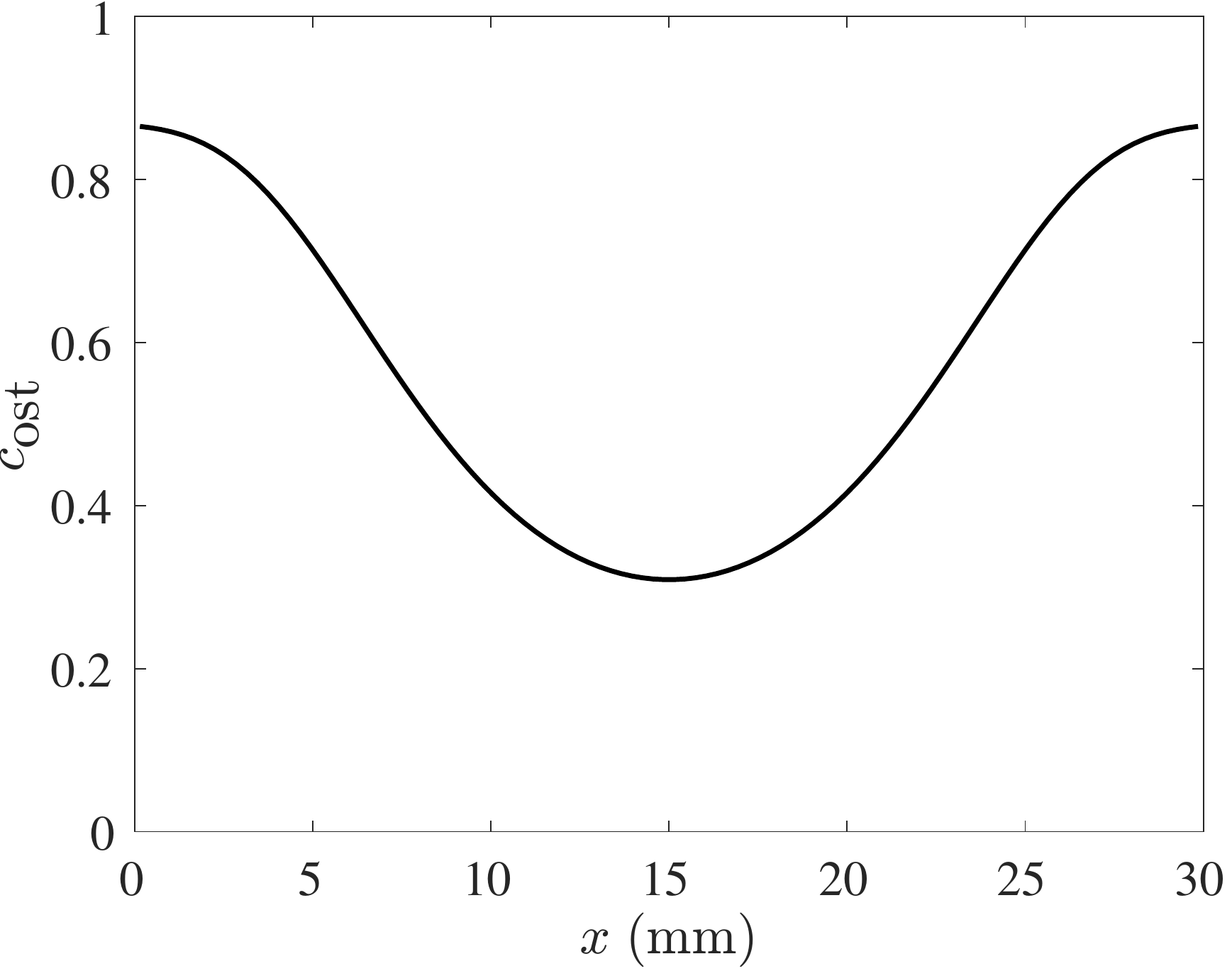}
\caption{Regenerated bone density (osteoblasts) $\cOst$  at $t=12\,\text{months}$.}
\end{subfigure}
\;
\begin{subfigure}[t]{0.3\textwidth}
\centering
\includegraphics[width=\textwidth]{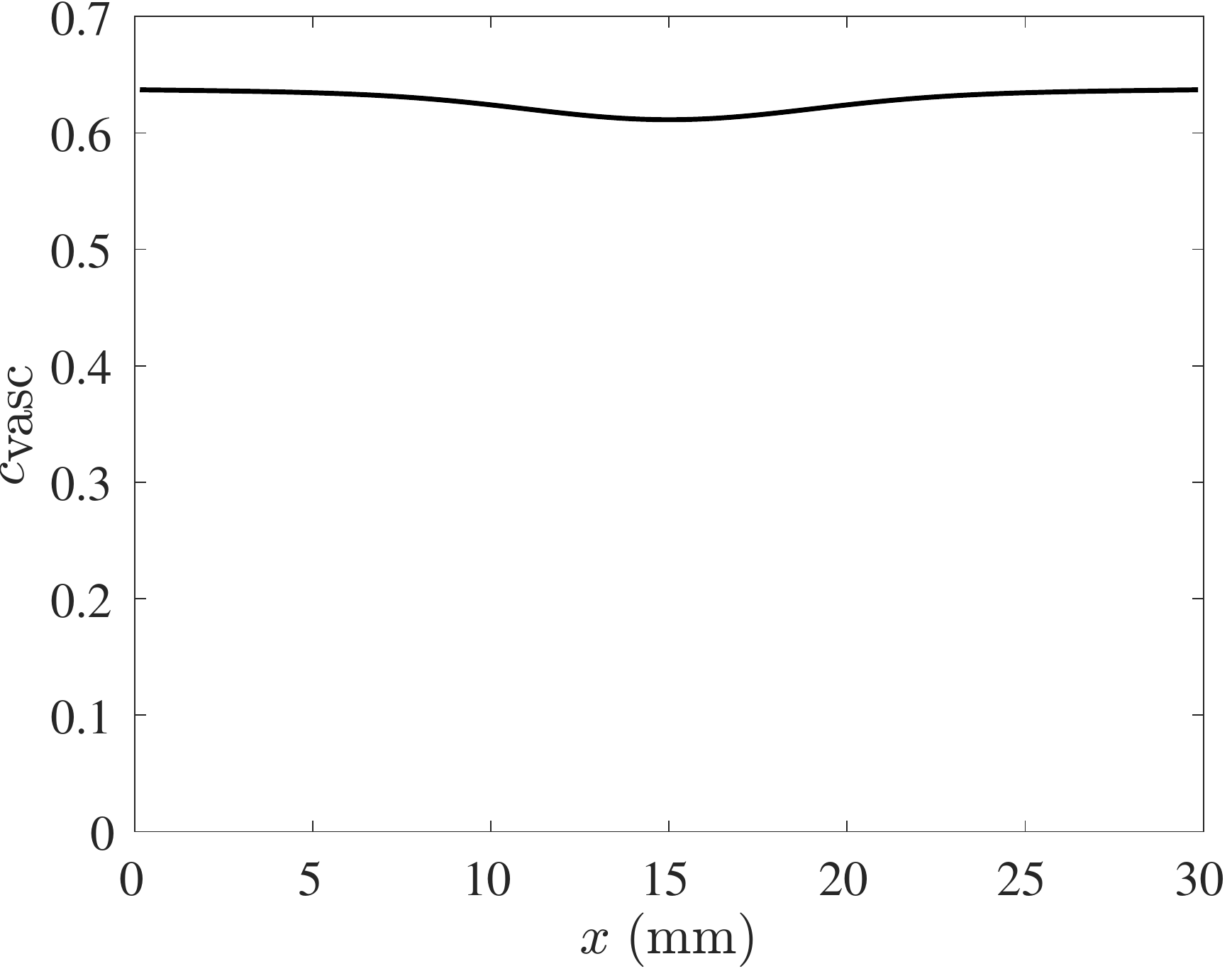}
\caption{Amount of regenerated vasculature $\cVasc$ at $t=12\,\text{months}$.}
\end{subfigure}
\;
\begin{subfigure}[t]{0.3\textwidth}
\centering
\includegraphics[width=\textwidth]{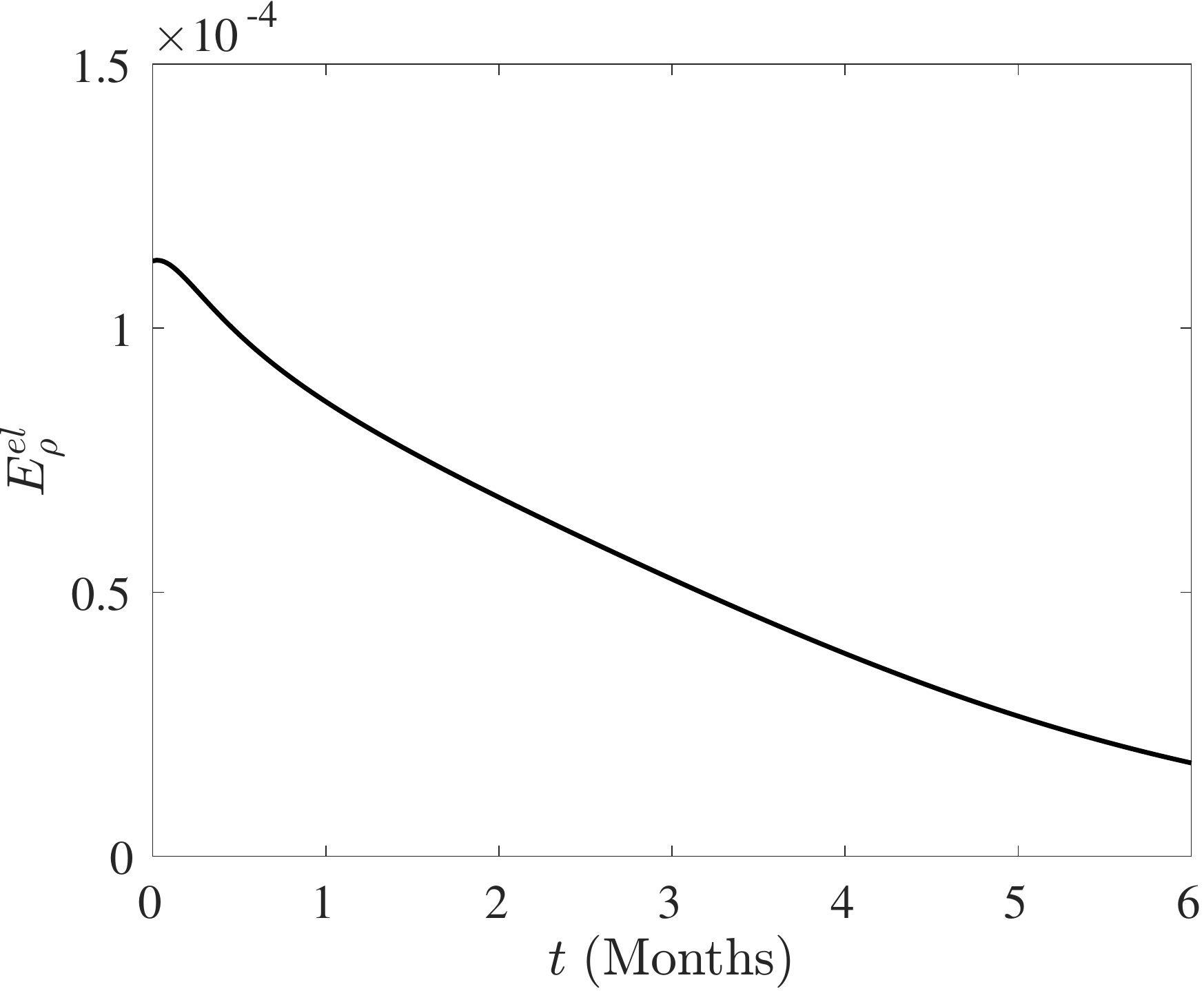}
\caption{Time evolution of the elastic energy in the scaffold/bone composite.}
\end{subfigure}
\caption{Outcome of the model for the default set of parameters given in Table \ref{tab:parameters} and a scaffold of constant density $\rho=0.13$. To be compared with \cite[Figure 2B, Scaffold only]{Cipitria:2015kp}.} \label{fig:default}
\end{figure}
\begin{figure}
\centering
\begin{subfigure}[t]{0.3\textwidth}
\centering
\includegraphics[width=\textwidth]{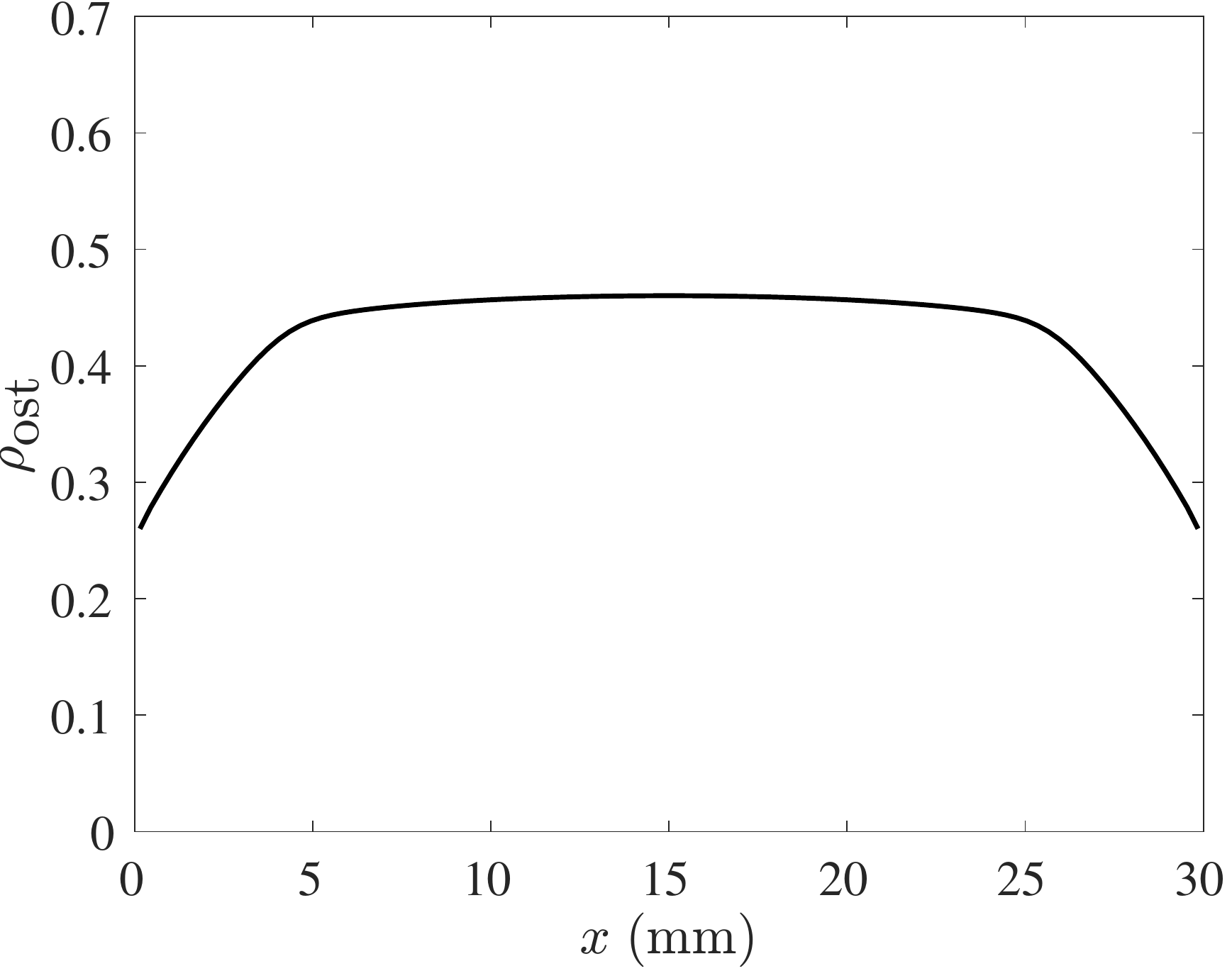}
\caption{Optimized scaffold density.}
\end{subfigure}
\;
\begin{subfigure}[t]{0.3\textwidth}
\centering
\includegraphics[width=\textwidth]{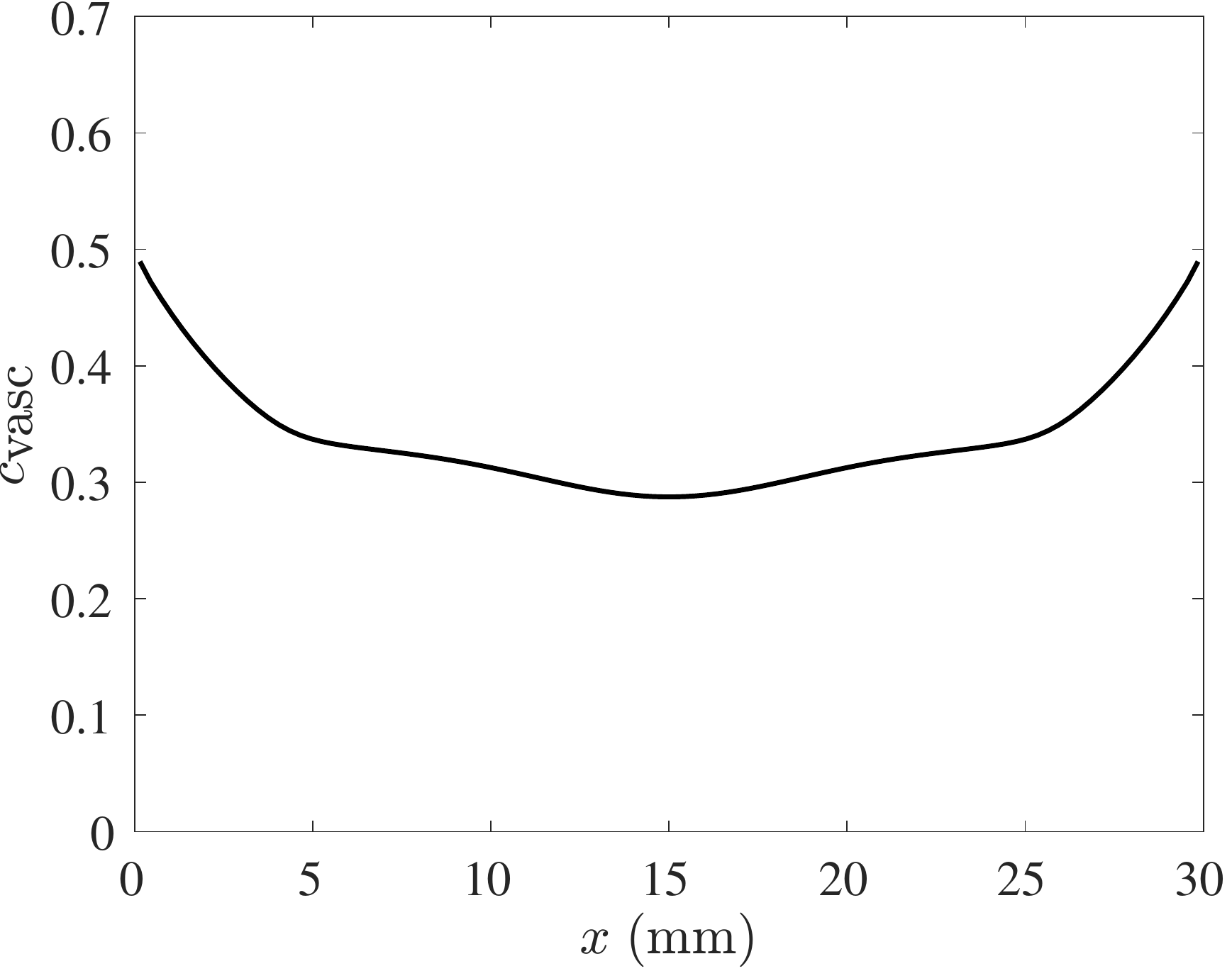}
\caption{Amount of regenerated vasculature at $t=12\,\text{months}$.}
\end{subfigure}
\;
\begin{subfigure}[t]{0.3\textwidth}
\centering
\includegraphics[width=\textwidth]{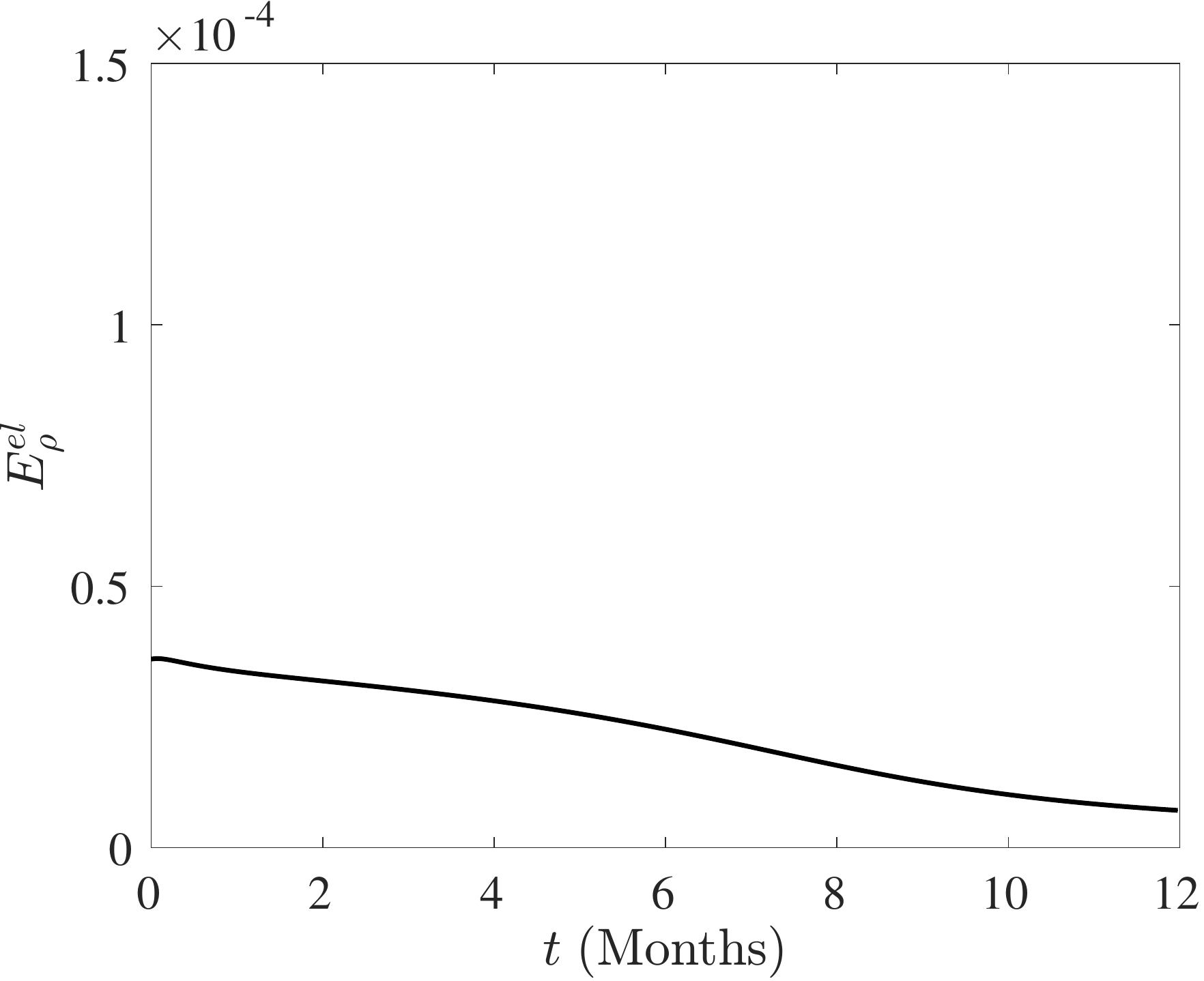}
\caption{Time evolution of the elastic energy in the scaffold/bone composite.}
\end{subfigure}
\caption{Optimal scaffold design for the default set of parameters given in Table \ref{tab:parameters}.} \label{fig:opt_default}
\end{figure}
The outcome of the model for the default set of parameters given in Table \ref{tab:parameters} can be seen in Figure \ref{fig:default}. One can clearly see that the distinct shape of regenerated bone bone density, with in-growth first from the proximal and distal end of the scaffold, is recovered in this model. The optimal scaffold design for the default set of parameters is displayed in Figure \ref{fig:opt_default}.

\begin{figure}
\centering
\begin{subfigure}[t]{0.3\textwidth}
\centering
\includegraphics[width=\textwidth]{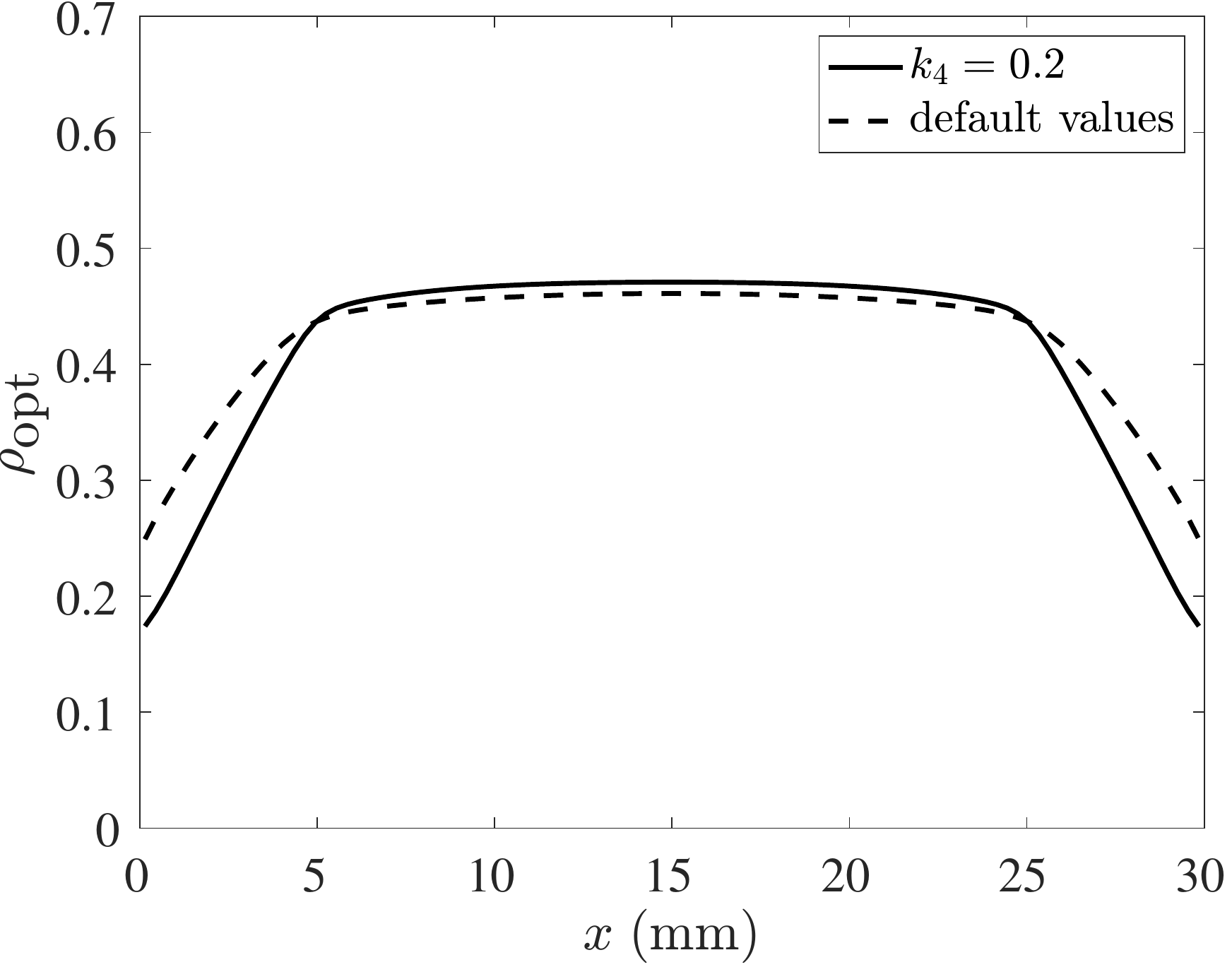}
\caption{Optimized scaffold density.}
\end{subfigure}
\;
\begin{subfigure}[t]{0.3\textwidth}
\centering
\includegraphics[width=\textwidth]{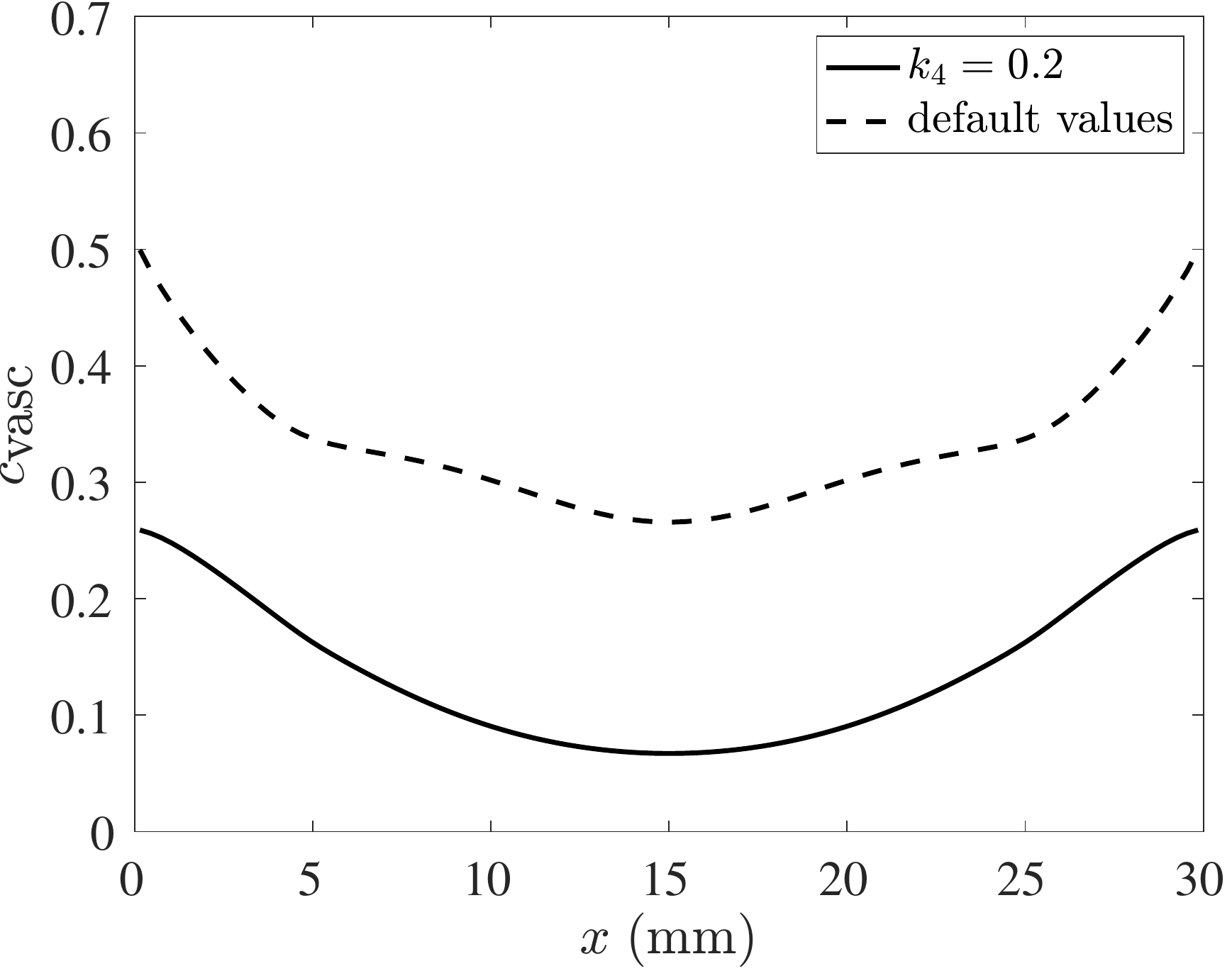}
\caption{Amount of regenerated vasculature at $t=12\,\text{months}$.}
\end{subfigure}
\;
\begin{subfigure}[t]{0.3\textwidth}
\centering
\includegraphics[width=\textwidth]{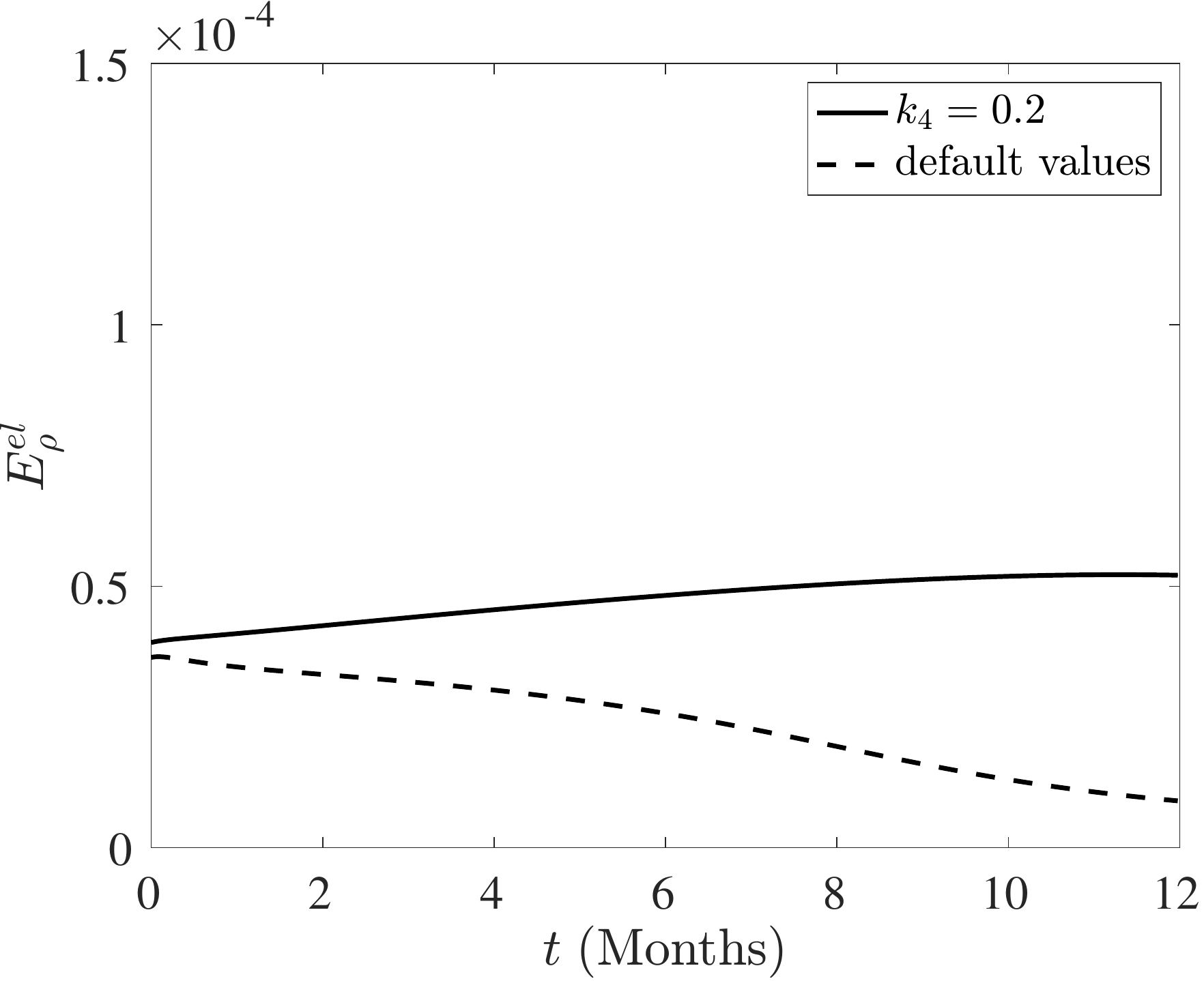}
\caption{Time evolution of the elastic energy in the scaffold/bone composite.}
\end{subfigure}
\caption{Optimal scaffold design for different values of $\genB$.} \label{fig:regrate}
\end{figure}

\begin{figure}
\centering
\begin{subfigure}[t]{0.3\textwidth}
\centering
\includegraphics[width=\textwidth]{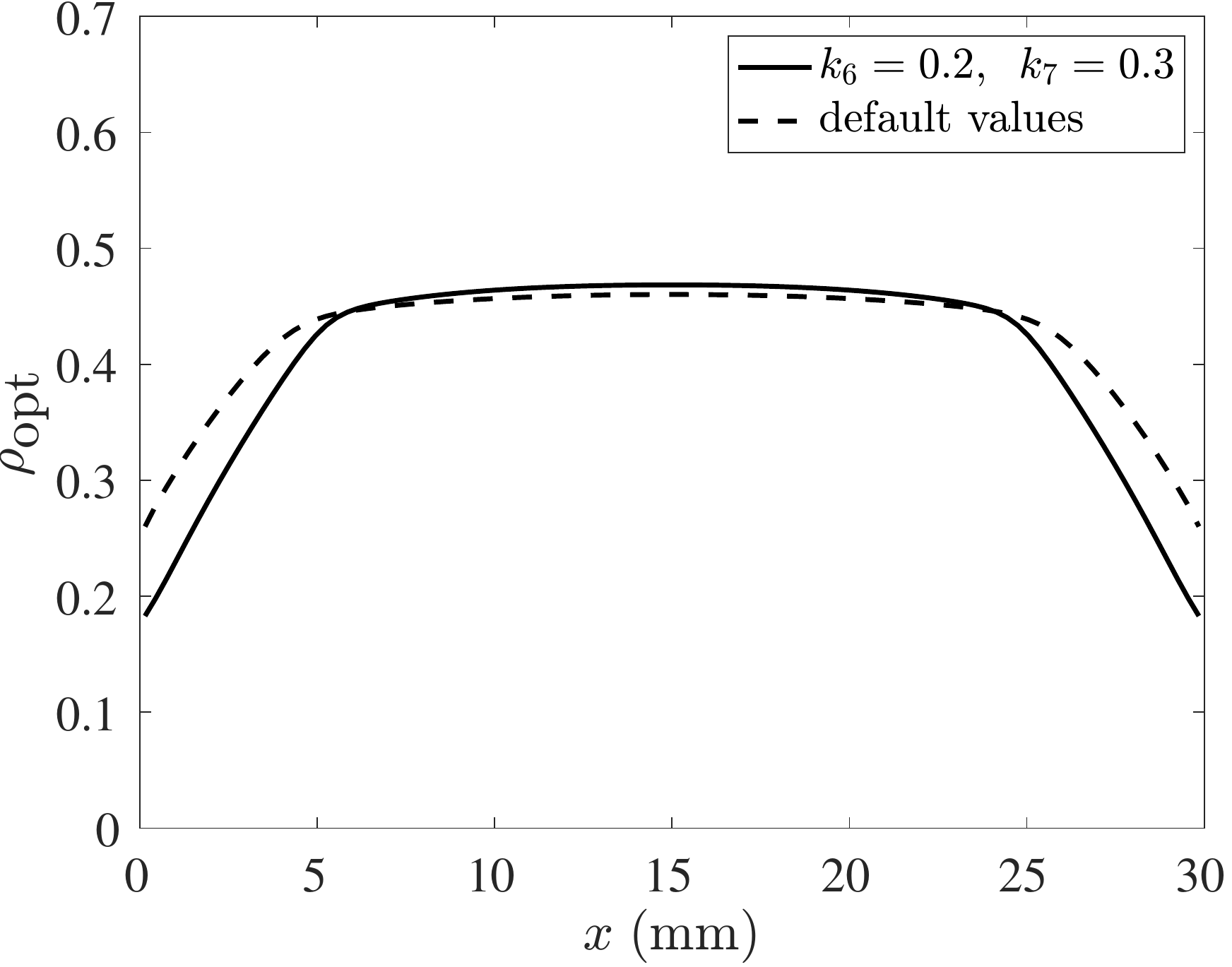}
\caption{Optimized scaffold density.}
\end{subfigure}
\;
\begin{subfigure}[t]{0.3\textwidth}
\centering
\includegraphics[width=\textwidth]{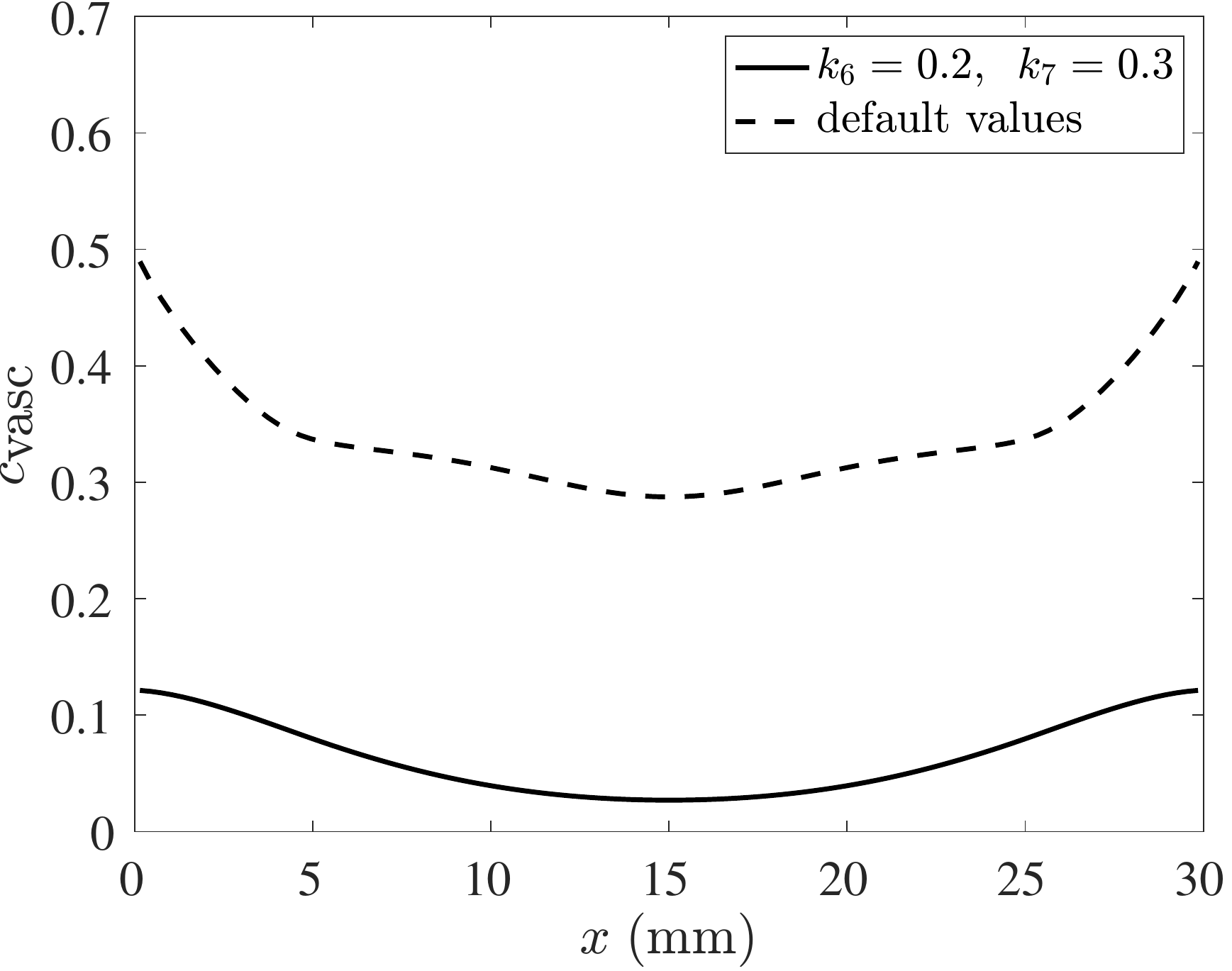}
\caption{Amount of regenerated vasculature at $t=12\,\text{months}$.}
\end{subfigure}
\;
\begin{subfigure}[t]{0.3\textwidth}
\centering
\includegraphics[width=\textwidth]{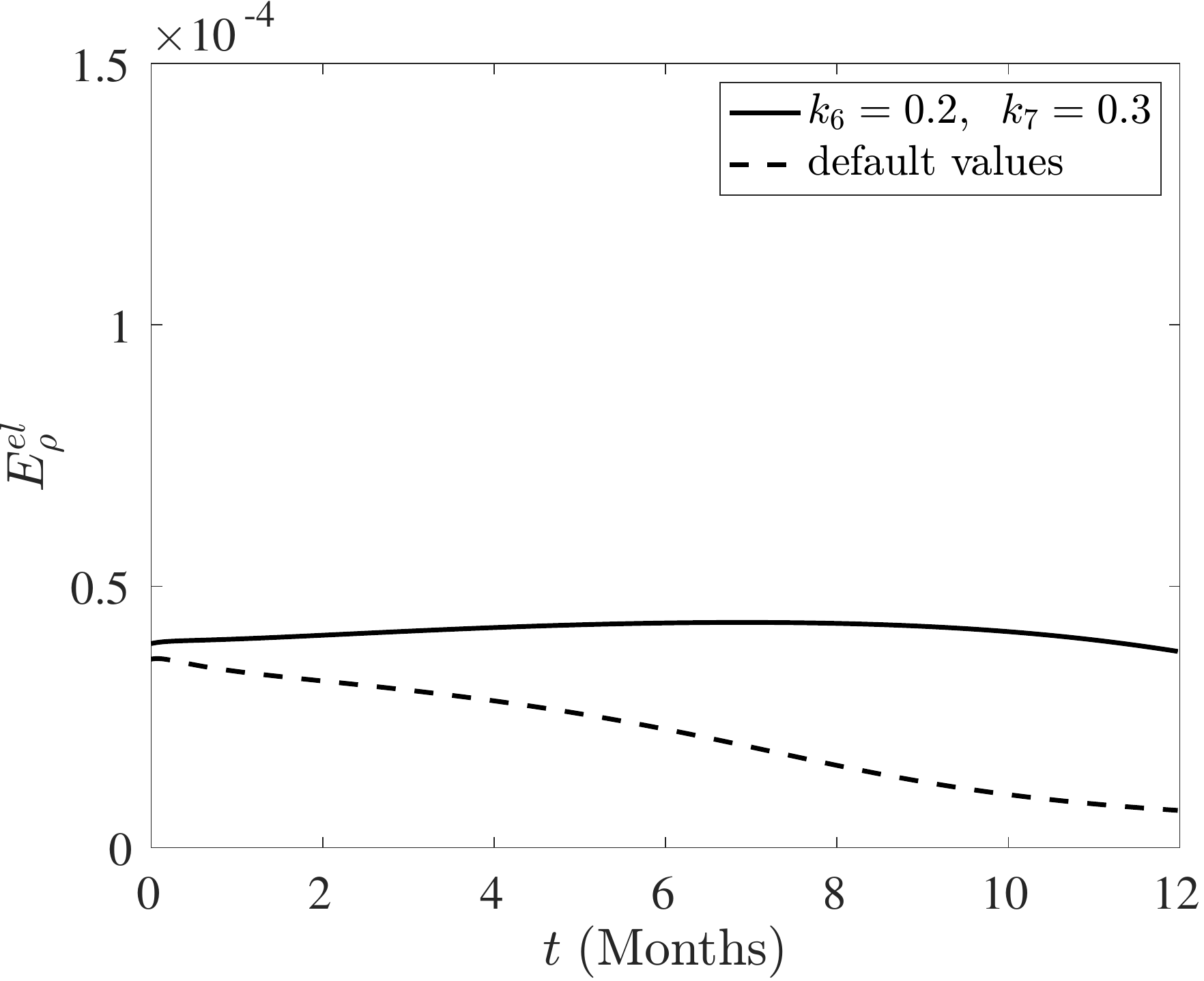}
\caption{Time evolution of the elastic energy in the scaffold/bone composite.}
\end{subfigure}
\caption{Optimal scaffold design for different values of $\genV$ and $\proV$.} \label{fig:vascrate}
\end{figure}

\begin{figure}
\centering
\begin{subfigure}[t]{0.3\textwidth}
\centering
\includegraphics[width=\textwidth]{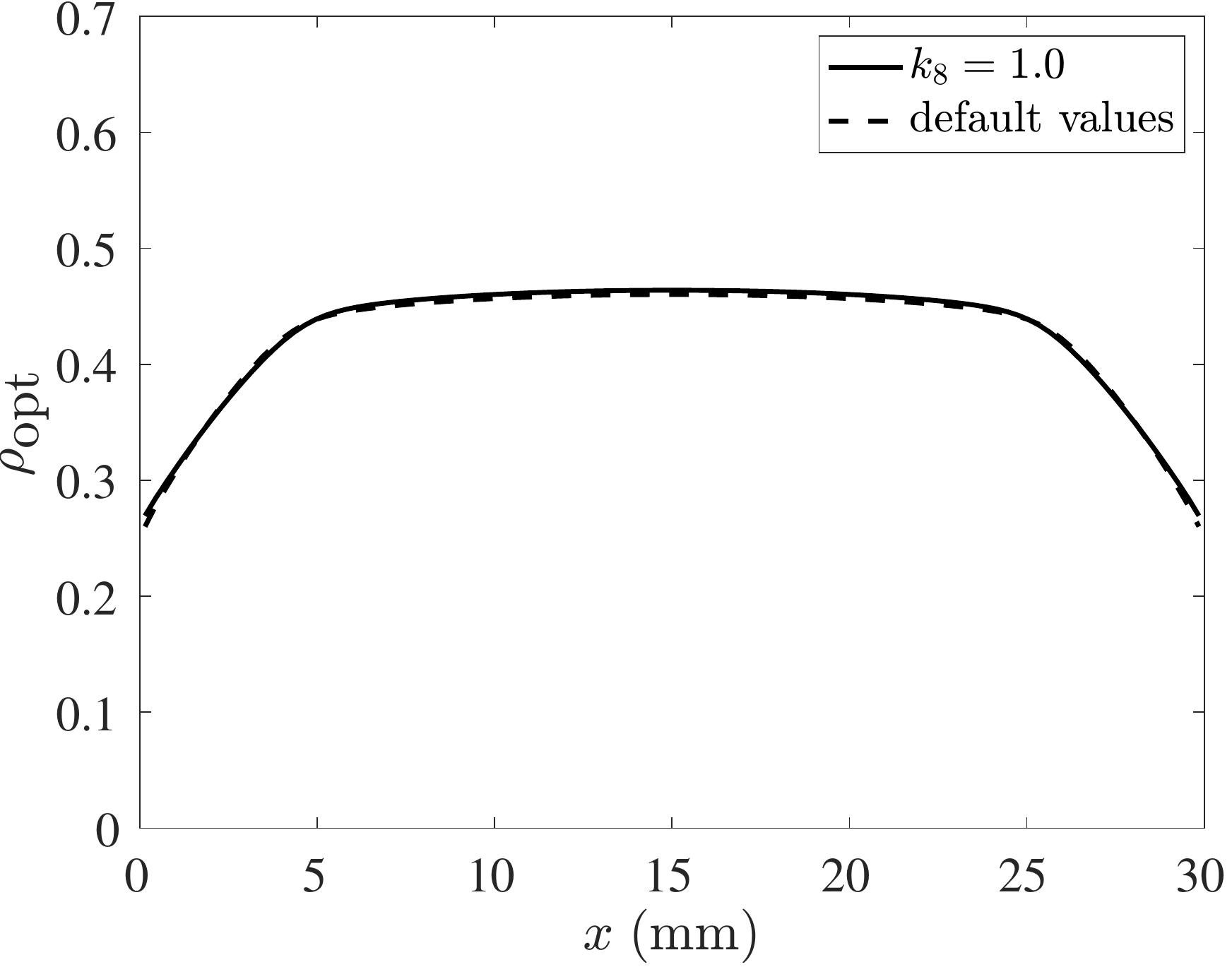}
\caption{Optimized scaffold density.}
\end{subfigure}
\;
\begin{subfigure}[t]{0.3\textwidth}
\centering
\includegraphics[width=\textwidth]{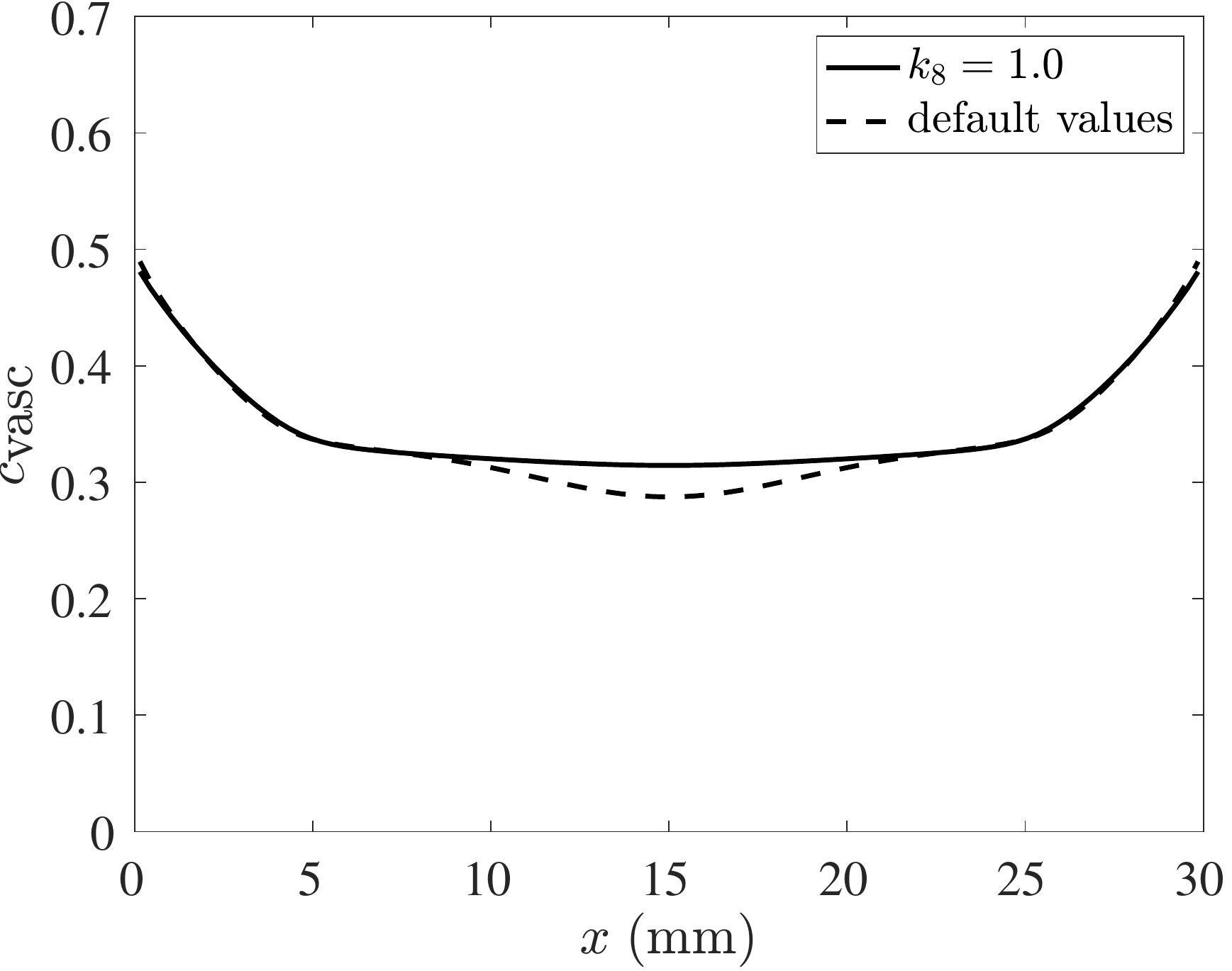}
\caption{Amount of regenerated vasculature at $t=12\,\text{months}$.}
\end{subfigure}
\;
\begin{subfigure}[t]{0.3\textwidth}
\centering
\includegraphics[width=\textwidth]{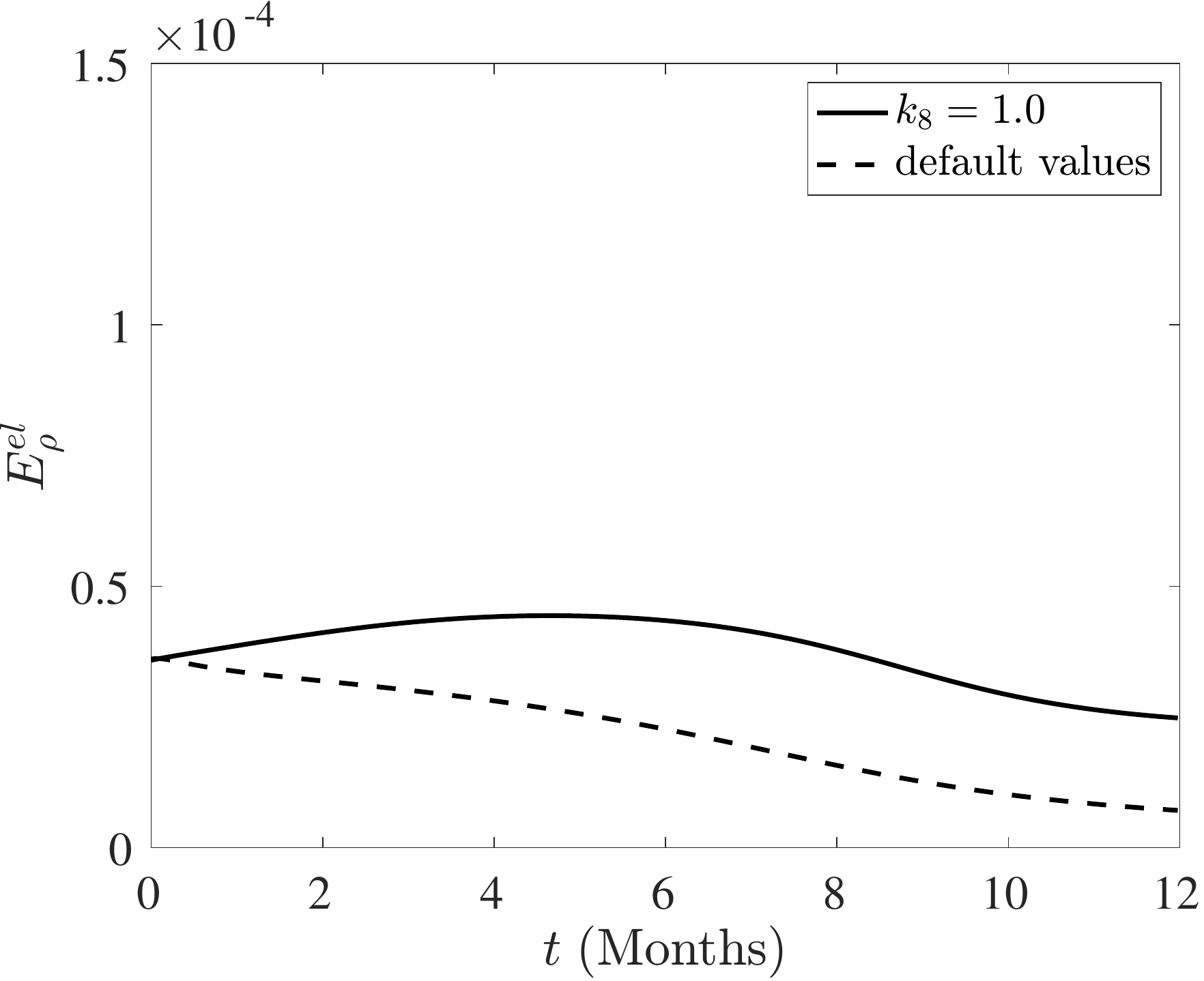}
\caption{Time evolution of the elastic energy in the scaffold/bone composite.}
\end{subfigure}
\caption{Optimal scaffold design for different values of $\relBone$.} \label{fig:bonedens}
\end{figure}

\paragraph{Experiment 1.} Varying the rate of regeneration. Figure \ref{fig:regrate} shows the optimal scaffolds for three different rates of bone regeneration $\genB$. All parameters other than $\genB$ are as given in Table \ref{tab:parameters}.

\paragraph{Experiment 2.} Varying the rate of vasculature formation. Figure \ref{fig:vascrate} shows the optimal scaffolds for three different rates for the formation of vasculature $\genV$, $\proV$. All parameters other than $\genV$, $\proV$ are as given in Table \ref{tab:parameters}.

\paragraph{Experiment 3.} Varying the relative stiffness of regenerated bone matrix/osteoblasts. Figure \ref{fig:bonedens} shows the optimal scaffolds for three different densities of regenerated bone matrix (here indicated by the relative elastic modulus $\relBone$). All parameters other than $\relBone$ are as given in Table \ref{tab:parameters}.

\paragraph{Summary.} Overall, one can note that impeded regeneration (as in experiments 1 and 2), the optimal scaffold is somewhat less dense at the proximal and distal ends of the defect (where the defect is adjacent to remaining healthy bone. This makes it easier for BGF to diffuse into the defect domain, thus accelerating bone regeneration. If the mechanical properties of regenerated bone are somehow compromised (due to, e.g., osteoporosis), our analysis yields no significant change in optimal scaffold architecture. The cost functional is still increased, however.

\bibliographystyle{alphabbr}
\bibliography{bones}

\end{document}